\documentclass[preprint]{revtex4-1}
 \usepackage{graphicx,psfrag,epsfig}

\newcommand{\beq}{\begin{equation}}
\newcommand{\eeq}{\end{equation}}

\newcommand{\bea}{\begin{eqnarray}}
\newcommand{\eea}{\end{eqnarray}}

\newcommand{\ea}{{\it et al.}}

\def\gsim{\mathrel{\vcenter{\hbox{$>$}\nointerlineskip\hbox{$\sim$}}}}
\begin{document}
\begin{flushright}
UMD-PP-10-006\\
March, 2010
\end{flushright}
\vspace{0.3in}
\title {\LARGE Electroweak Symmetry Breaking and Proton Decay in $SO(10)$ SUSY-GUT with TeV $W_R$}
\author{\bf P. S. Bhupal Dev and R. N. Mohapatra}
\affiliation{Maryland Center for Fundamental Physics,
Department of Physics, University of Maryland, College Park, MD
20742, USA}

\begin{abstract}
In a recent paper, we proposed a new class of supersymmetric
$SO(10)$ models for neutrino masses where the TeV scale
electroweak symmetry is $SU(2)_L\times SU(2)_R\times U(1)_{B-L}$
making the associated gauge bosons $W_R$ and $Z'$ accessible at
the Large Hadron Collider. We showed that there exists a
domain of Yukawa coupling parameters and symmetry breaking
patterns which give an excellent fit to all fermion masses
including neutrinos. In this sequel, we discuss an alternative
Yukawa pattern which also gives good fermion mass fit and then
study the predictions of both models for proton lifetime.
Consistency with current experimental lower limits on
proton life time require the squark masses of first two
generations to be larger than $\sim$ 1.2 TeV.
We also discuss how one can
have simultaneous breaking of both $SU(2)_R\times U(1)_{B-L}$ and
standard electroweak symmetries via radiative corrections.
\end{abstract}
\maketitle

\section{Introduction}
The nature of TeV scale new physics beyond the standard model (SM) is a
question of enormous interest as the Large Hadron Collider (LHC) is
poised to collect data in this energy range. Clearly, supersymmetry (especially the minimal
supersymmetric extension of the standard model (MSSM)) is one of
the prime candidates for this new physics since it not only solves the
gauge hierarchy problem, but also has a number of attractive
features such as the unification of gauge couplings at a high
scale, a potential dark matter candidate, etc.. An interesting
question along these lines has always been to see if any other new
physics can co-exist with TeV scale supersymmetry without
conflicting with coupling unification and dark matter, thereby
broadening the scope of LHC physics search.

A particularly appealing possibility is that weak
interactions conserve parity asymptotically~\cite{lr} with the associated
gauge group being  $SU(2)_L\times SU(2)_R\times U(1)_{B-L}$ so that the
resulting gauge bosons $W_R$ and $Z'$ are at the TeV scale co-existing
with supersymmetry. The case for $SU(2)_L\times SU(2)_R\times U(1)_{B-L}$
becomes more compelling when the SM or MSSM
are extended to understand small neutrino masses via the seesaw
mechanism~\cite{seesaw}. As a generic possibility, this scenario is quite
consistent with current low energy observations.
Whether a TeV Scale $SU(2)_R$ symmetry is compatible with supersymmetric
coupling unification
 has been extensively investigated in literature~\cite{many,desh}.
With a few exceptions~\cite{desh}, it seems very hard to reconcile
this possibility with the observed value of $\sin^2{\theta_W}$. In
a recent paper~\cite{model}, we pointed out a new supersymmetric $SO(10)$
scenario where the presence of a vector like electroweak singlet and
color triplet Higgs multiplet (which is part of the {\bf 45}
representation in $SO(10)$) in addition to two bidoublets and two
right handed doublets of the left-right electroweak group at the
TeV scale leads to gauge coupling unification with TeV scale right
handed $W_R$ and $Z'$ bosons . This model is different from other
such scenarios considered in the literature~\cite{desh} in that
quark masses and mixing arise in a simple manner. The neutrino
masses arise
 out of an inverse seesaw mechanism~\cite{MV} and was
shown~\cite{model} to have interesting phenomenological
consequences like leptonic non-unitarity, leptonic $CP$-violation,
lepton flavor violation, etc. which may be testable in near
future. This fit to the fermion masses defines one class of $SO(10)$
models with TeV scale $W_R$ which we call model (A).

In this paper, several new results for these $SO(10)$ models are
presented: (i) we present an alternative fit to fermion masses,
which we call model (B); (ii) we discuss the constraints of proton
decay for both fermion mass fits -- the one in Ref.~\cite{model} and the
new one discussed in this paper; (iii) we also show how both $B-L$
and electroweak symmetries can be broken radiatively in these
models.

Strength of proton decay has been studied extensively in the
context of many  supersymmetric grand unified theories (SUSY GUTs) (see
Ref.~\cite{review} for recent reviews). Although there is no evidence
for proton decay till now, current experimental lower bounds on
the partial lifetimes of various proton decay modes tend to put
severe constraints on these models e.g. they have now ruled out
the simplest versions of SUSY $SU(5)$ and suggest possible
modifications of such models~\cite{bajc}. They also constrain the
choices of Higgs multiplets that can be used for model building
with $SO(10)$ group~\cite{babu}.

In the models we are discussing here, due to the fact that all the
Yukawa couplings responsible for proton decay are constrained by
the fermion mass fits, it is possible to estimate the partial life
times for the various modes as functions of the squark masses and
for reasonable squark masses of the first two generations, and for
model (A), we get upper bounds on various proton decay channels.
There are no such bounds in the second case (model (B)). We find that within a
reasonable set of assumptions, all our predicted upper bounds for
model (A)  are consistent with the current experimental bounds and
some of the modes  may be accessible to the next generation proton
decay experiments with megaton size detectors.

We also discuss the constraints imposed by radiative breaking of both
$SU(2)_R\times U(1)_{B-L}$ and the SM gauge symmetries
via radiative corrections. The idea is to start with soft mass squares at
the Planck or GUT scale and extrapolate the masses to the weak scale to
see if the $SU(2)_R\times U(1)_{B-L}$ symmetry breaks at the TeV scale.
We then note that this breaking introduces via $D$-terms a breaking of the
SM gauge symmetry to $U(1)_{\rm em}$.

We also discuss the generalization of this model to include $R$-parity
breaking and its implications on proton decay.

This paper is organized as follows: in Sec. II, we review the basic
structure of our model and the gauge symmetry breaking. In Sec. III, we
review the fermion mass fit for model (A) already discussed in 
Ref.~\cite{model}.
In Sec. IV, we present a new fermion mass fit and define it as model (B).
Sec. V describes the radiative electroweak symmetry breaking (EWSB) in this type of models.  In Sec. VI, we
discuss the proton decay in both these models.
In Sec. VII, we comment on the effect of $R$-parity breaking terms in the
superpotential on proton decay. The results are summarized in Sec. VIII.
In Appendix A, we present the renormalization group equations (RGEs) for soft
SUSY-breaking masses in our supersymmetric left-right (SUSYLR) model.
In Appendix B, we derive the anomalous dimensions of the
dimension-5 proton decay operators in our model. In Appendix C, we list the
hadronic form factors used in our proton decay calculations.
\section{A brief overview of the model}
As in the usual $SO(10)$ models, the three generations of quark
and lepton fields are assigned to three {\bf 16} dim. spinor
representations. In addition, we add three $SO(10)$ singlet matter
fields to implement the inverse seesaw mechanism. The $B-L$ gauge
symmetry is broken at the TeV scale by {\bf 16}-Higgs fields
(denoted by $\psi_H$), whereas the rest of the gauge symmetry is
broken at $\sim 10^{16}$ GeV by {\bf 54} and {\bf 45}- fields
(denoted by $E$ and $A_a$ respectively).  We require
two {\bf 45}-Higgs fields ($a=1,2$), one for symmetry breaking and
the other to give rise to the vector-like color triplets at the
TeV scale. The SM symmetry is broken by two {\bf 10}-Higgs fields
(denoted by $H_a$). We note that the field content of our
model is found in many string models after compactification e.g.
fermionic compactification models~\cite{raby} and it may therefore
be easier to embed this GUT model into strings.

 The distinguishing feature of our model is that the GUT symmetry
breaks down to the left-right symmetric gauge group $SU(3)_c\times
SU(2)_L\times SU(2)_R\times U(1)_{B-L}$ without parity
($D$-parity). The $D$-parity is broken at the GUT scale by the vacuum
expectation value (VEV) of the {\bf 45}-Higgs field. A consequence of $D$-parity breaking
is that only the right-handed (RH) doublets from {\bf 16}-Higgs fields survive
below the GUT scale.  An interesting feature of this class of
models~\cite{model} is that if we have two RH Higgs fields
[$\chi^c, \bar{\chi}^c~ (1,1,2,\pm 1)$] , two bi-doublet fields
[$\Phi(1,2,2,0)$] (all color singlets) and a vector-like color
triplet but $SU(2)_L\times SU(2)_R$ singlet field
[$\delta\left(3,1,1,\frac{4}{3}\right)$+c.c.] at the TeV scale,
the gauge couplings unify around $10^{16}$ GeV. The bidoublet
fields arise from {\bf 10}-Higgs at the GUT scale and the
vector-like color triplet fields arise from the {\bf 45}-Higgs
field. This is therefore a new class of $SO(10)$ SUSY-GUT theories
with TeV scale $W_R$ and $Z'$ bosons which can be accessible at
the LHC.

We consider the symmetry breaking chain
\begin{eqnarray}
    SO(10)\stackrel{M_G}{\longrightarrow}
\mathbf 3_c \mathbf 2_L \mathbf 2_R \mathbf 1_{B-L}
\stackrel{M_R}{\longrightarrow}
\mathbf 3_c \mathbf 2_L \mathbf 1_Y ({\rm MSSM})
\stackrel{M_{\rm SUSY}}{\longrightarrow}
\mathbf 3_c \mathbf 2_L \mathbf 1_Y ({\rm SM})
\stackrel{M_Z}{\longrightarrow}
\mathbf 3_c \mathbf 1_Q
\label{eq:chain}
\end{eqnarray}
where, as an example of our notation, ${\bf 3}_c$ means $SU(3)_c$. As shown in
Appendix A of Ref.~\cite{model}, for consistency, we need at
least two ${\bf 45}$ and one ${\bf 54}$ representations of the Higgs fields
to break the $SO(10)$ gauge group into the SUSYLR gauge group,
$SU(3)_c \times SU(2)_L\times SU(2)_R\times U(1)_{B-L}$, at the scale
$M_G\simeq
4\times 10^{16}$ GeV.
Note that to have realistic fermion masses and mixing, we need at
least two $SU(2)$ bi-doublets of the ${\bf 10}$ Higgs
representation to break the $SU(2)_L\times U(1)_Y$ gauge group of
the SM to $U(1)_Q$ at the weak scale $M_Z$. With this minimal set
of Higgs fields, we were able to attain not only gauge coupling
unification but also the desired fermion masses and mixing at the
GUT scale~\cite{model}. Incidentally, since our gauge group
above TeV scale is different from MSSM, we needed to extrapolate
fermion masses using the left-right group (see Appendix B of
Ref.~\cite{model}) which has certain distinguishing features in the running
behavior, in contrast to the MSSM gauge group.

The superpotential for the model consists of several parts:
\begin{equation}
W~=~W_{SB}+~W_m~+W^\prime
\end{equation}
where $W_{SB}$ is responsible for $SO(10)$ GUT symmetry breaking,
doublet triplet splitting and the remnant sub-GUT scale
multiplets; $W_m$ is the Yukawa superpotential responsible for
fermion masses and mixing; $W^\prime$ involves the $R$-parity
violating terms. When we impose an additional matter parity
symmetry under which $\psi_\alpha \to -\psi_\alpha,~ S_\alpha \to
-S_\alpha$, and all other fields even, as was assumed
in Ref.~\cite{model}, we get $W^\prime~=0$, i.e. all $R$-parity
violating terms are absent in the superpotential and the model has
a stable dark matter~\cite{arina}.  We discuss the effects of
nonzero $W^\prime$ in a subsequent section where we show that even
after including arbitrary $R$-parity violating terms (i.e. giving
up matter parity assumption), the model does satisfy proton life
time bounds since $W^\prime$ conserves baryon number and after
$B-L$ breaking leads to a highly suppressed amplitude for proton
decay. This feature is characteristic only of $SO(10)$ models with
low $B-L$ breaking.

The Yukawa superpotential is given by
\begin{eqnarray}
    W_m = h_{aij}{\bf 16}_i{\bf 16}_j{\bf 10}_{H_a}+~\frac{f_{aij}}{M^2}
    {\bf 16}_i{\bf 16}_j{\bf 10}_{H_a}{\bf 45}_H{\bf 45}'_H
    \label{eq:yuksup}
\end{eqnarray}
where the first term is the usual Yukawa coupling term, while the
second term is a higher-dimensional term whose completely
antisymmetric combination acts as an effective ${\bf 126}_H$
operator, thus giving rise to a realistic fermion mass spectrum at
the GUT scale. We define this as our model (A).

 The superpotential $W_{SB}$ was discussed in detail in Ref.~\cite{model} 
 where it was noted
that the following components of the ${\bf 54}$ and ${\bf 45}$
Higgs fields acquire VEV and leave the left-right subgroup
unbroken:
\begin{eqnarray}
&&\langle{\bf 54}\rangle = {\rm diag}\left(a,a,a,a,a,a,-\frac{3}{2}a,
-\frac{3}{2}a,-\frac{3}{2}a,-\frac{3}{2}a\right);\nonumber\\
&&\langle{\bf 45}\rangle_{12} = \langle{\bf 45}\rangle_{34} =
\langle{\bf 45}\rangle_{56} = b.
\end{eqnarray}
\section{Fermion masses in model (A)}
The model discussed in Ref.~\cite{model} is defined by the VEV pattern of the
bi-doublets:
\begin{eqnarray}
    \langle \Phi_1\rangle=\left(\begin{array}{cc} \kappa_d & 0\\ 0 & 0
    \end{array}\right),~ ~
    \langle\Phi_2\rangle =\left(\begin{array}{cc} 0 & 0\\ 0 & \kappa_u
    \end{array}\right)
    \label{eq:veva}
\end{eqnarray}
We define the ratio of the VEVs
as $\tan\beta\equiv \frac{\kappa_u}{\kappa_d}$ as in MSSM. Then the fermion
mass matrices at the GUT-scale are given by
\begin{eqnarray}
    M_u &=& \tilde{h}_u+\tilde{f},\nonumber\\
    M_d &=& \tilde{h}_d+\tilde{f},\nonumber\\
    M_e &=& \tilde{h}_d-3\tilde{f},\nonumber\\
    M_{\nu_D} &=& \tilde{h}_u-3\tilde{f}.
    \label{eq:mass}
\end{eqnarray}
where in the notation of Ref.~\cite{model}, $\tilde{h}_{u,d}\equiv
\kappa_{u,d}h_{u,d}$. The contribution from the effective $\mathbf {126}_H$
operator is assumed to be the same for both up and down sectors, i.e.
$\tilde{f} = \kappa_u f_u = \kappa_d f_d$; as a result,
we have the relation $f_d=f_u\tan\beta$. Also note that the
factor $-3$ between the quark and lepton sector is due to the same
${\bf 126}$ operator.
Using the renormalization group analysis for the fermion masses and mixing in
the SUSYLR
model (see Appendix B of Ref.~\cite{model}), we obtain the
GUT-scale fermion masses starting from the experimentally known
values at the weak scale. Using these mass values, we obtain a fit
for the coupling matrices at the GUT scale
defined in Eq.~(\ref{eq:mass}). Here we give the results in a down quark mass diagonal basis for two cases: \\
(a) $\tan{\beta}_{\rm MSSM}=10$: In this case, the GUT-scale values of the
charged fermion masses are found to be
\begin{eqnarray}
    m_u = 0.0017~{\rm GeV},~ ~ m_c = 0.1908~{\rm GeV},~ ~ m_t = 77.7~
    {\rm GeV},\nonumber\\
    m_d = 0.0013~{\rm GeV},~ ~ m_s = 0.0263~{\rm GeV},~ ~ m_b = 1.7001~{
    \rm GeV},\nonumber\\
    m_e = 0.0004~{\rm GeV},~ ~m_\mu = 0.0910~{\rm GeV},~ ~m_\tau = 1.7061~
    {\rm GeV}
    \label{eq:m10}
\end{eqnarray}
and $\tan{\beta}_{\rm GUT}=7$. Note that the GUT-scale fermion
masses quoted here are slightly different from those given in
Ref.~\cite{model} because, in this case, we have set the $S\Phi\Phi$ coupling
$\mu_{\Phi}=0$ (of Ref.~\cite{model}) assuming $R$-parity conservation.
With these mass eigenvalues, we find a fit for the GUT-scale
couplings of the form:
\begin{eqnarray}
    f_u &=& {\rm diag}~\left(1.26\times 10^{-6},-0.0001,-9.48\times 10^{-6}
    \right),~ ~ f_d = f_u\tan{\beta}_{\rm GUT},\nonumber\\
    h_d &=& {\rm diag}~\left(4.86\times 10^{-5},0.0019,0.0752\right),\nonumber\\
    h_u &=& \left(\begin{array}{ccc}
        7.46\times 10^{-5} & 0.0002-6.51\times 10^{-5}i & 0.0002-0.0028i\\
        0.0002+6.51\times 10^{-5}i & 0.0015 & 0.0118+1.26\times 10^{-6}i\\
        0.0002+0.0028i & 0.0118-1.26\times 10^{-6}i & 0.4908
    \end{array}\right)
    \label{eq:hud10}
\end{eqnarray}
Note that for simplicity we have chosen the $f$-couplings to be diagonal.
Our fit does not allow the off-diagonal components to be too different from
zero.\\\\
(b) $\tan{\beta}_{\rm MSSM} = 30$: In this case, the GUT-scale values of the
charged fermion masses are found to be
\begin{eqnarray}
    m_u = 0.0121~{\rm GeV},~ ~ m_c = 0.3269~{\rm GeV},~ ~ m_t = 120.53~
    {\rm GeV},\nonumber\\
    m_d = 0.0014~{\rm GeV},~ ~ m_s = 0.0277~{\rm GeV},~ ~ m_b = 2.7958~{
    \rm GeV},\nonumber\\
    m_e = 0.0006~{\rm GeV},~ ~m_\mu = 0.1266~{\rm GeV},~ ~m_\tau = 2.7737~
    {\rm GeV}
    \label{eq:m30}
\end{eqnarray}
and $\tan{\beta}_{\rm GUT}=20$.
With these mass eigenvalues, we obtain a fit for the
couplings of the following form:
\begin{eqnarray}
    f_u &=& {\rm diag}~\left(1.5\times 10^{-6},-0.0002,4.2\times 10^{-5}
    \right),~ ~ f_d = f_u\tan{\beta}_{\rm GUT},\nonumber\\
    h_d &=& {\rm diag}~(0.0002,0.0078,0.4163),\nonumber\\
    h_u &=& \left(\begin{array}{ccc}
        0.0002 & 0.0003-0.0001i & -0.0008-0.0081i\\
        0.0002+0.0001i & 0.0029 & 0.0144+0.0002i\\
        -0.0008+0.0081i & 0.0144-0.0002i & 0.9145
\end{array}\right)
    \label{eq:hud30}
\end{eqnarray}
We note that in this model, larger values of $\tan{\beta}~(>30)$
are not allowed. This can be seen analytically from the form of
the RGEs given in Appendix B of Ref.~\cite{model} where it is
clear that the up-quark sector masses will increase rapidly at
high energies for large $\tan{\beta}$ and the same effect is
induced in the down-quark sector which makes the Yukawa terms
dominant over the gauge terms. This makes all the quark masses to
run up to unacceptably large values at the GUT-scale. We believe
this is a general feature of low-scale SUSYLR models, in contrast
to MSSM case~\cite{antusch}.
\section{A new fermion mass fit: Model (B)}
In this section, we consider an alternative mass fit within the
$SO(10)$ models with low scale $B-L$. It follows from a recent
ansatz~\cite{mimura} that in generic $SO(10)$ models which do not
use type I seesaw to fit neutrino masses, an alternative fit to
fermion masses is possible using the idea~\cite{mimura} that one
has a rank one {\bf 10}-Higgs Yukawa coupling matrix which
dominates the fermion masses while other couplings introduce small
corrections; the third generation masses arise from the dominant
rank one coupling matrix with smaller {\bf 126} and second {\bf
10} couplings generating the CKM mixing as well as the second and
the first generation fermion masses. This idea can be applied to
our case since, the neutrino mass is given by the inverse seesaw
formula which involves an additional matrix $\mu$. The main
difference of model (B) as compared to model (A) resides in the
VEV pattern of the two Higgs bidoublets i.e. in model (B), we
have
\begin{eqnarray}
    \langle \Phi_1\rangle = \left(\begin{array}{cc}
        \kappa_d & 0\\ 0 & \kappa_u
    \end{array}\right),~ ~
    \langle \Phi_2\rangle = \left(\begin{array}{cc}
        \kappa'_d & 0\\ 0 & \kappa'_u
    \end{array}\right)
    \label{eq:vevmodb}
\end{eqnarray}
with $v_{\rm wk}/\sqrt{2} =
\sqrt{\kappa_u^2+\kappa_d^2+\kappa_u'^2+\kappa_d'^2}$.
Also we must have $\frac{\kappa_u}{\kappa_d}\neq \frac{\kappa'_u}{\kappa'_d}$
in order to get right fermion mixing pattern.
In the limit $\kappa_u\gg \kappa'_u$,
the RG analysis of model (A) can be applied to this case
to generate fermion masses at the GUT scale as well as the
symmetry breaking pattern via radiative corrections.

The resulting fermion mass formulae in terms of the
appropriately redefined Yukawa couplings are given as follows~\cite{mimura1}:
\begin{eqnarray}
    M_u &=& \tilde{h}+r_2\tilde{f}+r_3\tilde{h}',\nonumber\\
    M_d &=& r_1(\tilde{h}+\tilde{f}+\tilde{h}'),\nonumber\\
    M_l &=& r_1(\tilde{h}-3\tilde{f}+c_e\tilde{h}'),\nonumber\\
    M_{\nu_D} &=& \tilde{h}-3\tilde{f}+c_\nu \tilde{h}'
\end{eqnarray}
where
\begin{eqnarray}
&&  \tilde{h} = \kappa_u h,~ ~ \tilde{f} = \frac{\kappa_u\kappa'_d}
    {\kappa_d}f,~ ~ \tilde{h}' = \frac{\kappa_u\kappa'_d}{\kappa_d}h',
    \nonumber\\
&&  r_1 = \frac{\kappa_d}{\kappa_u},~ ~ r_2 = r_3 = \frac{\kappa_d
    \kappa'_u}{\kappa_u\kappa'_d}.
\end{eqnarray}
As in the case of model (A), the $f$ coupling above represents the
effective {\bf 126} coupling arising from the $\psi\psi A_1A_2H_2$
term in the superpotential and $h'$ arises from a coupling of the
form $\psi\psi H_2 X$ (with a nonzero VEV for the additional singlet field $X$).
Note that if there is an additional $Z_2 $ symmetry under which
$H_2, A_2, X$ are odd and all other fields are even, one can have
a superpotential with only the $h,f,h'$ type contributions as
given above, to the fermion mass formulae. In our case with two
Higgs bi-doublets, $c_e=1$ and $c_\nu = r_3$. With the GUT-scale
mass eigenvalues obtained earlier, we obtain a fit for these couplings
as follows:\\
(a) $\tan{\beta}_{\rm MSSM} = 10$:
\begin{eqnarray}
    && \kappa_u = 173.2~{\rm GeV},~ ~r_1 = 0.0218,~ ~ r_2 = 0.14,\nonumber\\
    && h = {\rm diag}~(0,0,0.45),\nonumber\\
    && f = \left(\begin{array}{ccc}
        0 & -0.0006 & 0.0019\\
        -0.0006 & 0.0115 & 0.0101\\
        0.0019 & 0.0101 & 0.0001
    \end{array}\right),~ ~
    h' = i\left(\begin{array}{ccc}
    0 & -0.0022 & 0.0005\\
    0.0022 & 0 & 0.0181\\
    -0.0005 & -0.0181 & 0
\end{array}\right)
\label{eq:byuk10}
\end{eqnarray}
(b) $\tan{\beta}_{\rm MSSM} = 30$:
\begin{eqnarray}
    && \kappa_u = 172.4~{\rm GeV}, r_1 = 0.0231,~ ~ r_2 = 0.21,\nonumber\\
 &&   h = {\rm diag}~(0,0,0.70),\nonumber\\
    && f = \left(\begin{array}{ccc}
        0 & -0.0016 & 0.0062\\
        -0.0016 & 0.0140 & 0.0111\\
        0.0062 & 0.0111 & 0.0019
    \end{array}\right)
\label{eq:byuk30}
\end{eqnarray}
and $h'$ same as in case (a). It may be noted here that in both the cases, 
all the fermion mass values predicted using the couplings above agree with those obtained from the RGEs
within the experimental uncertainty, the only exception being the up-quark mass 
in case (a), where the our predicted value is about 4 times larger. 
Note however that in  our discussion,
we have not included contributions from threshold
corrections or higher dimensional operators. Those contributions can
generally be of order MeVs when their couplings are chosen appropriately, in which case,
they will not affect the second and third generation masses
but could easily bring the up quark mass into agreement
with RGE predictions.

With the Yukawa couplings completely fixed in our model, we can
analyze the predictions for the proton decay rate. But before
doing so, we discuss the details of the electroweak symmetry
breaking in this model which was not done in the original
paper~\cite{model}. This discussion applies to both models (A) and
(B).
\section{Symmetry breaking by radiative corrections}
In this section, we propose a way to break both the $SU(2)_R\times
U(1)_{B-L}$ as well as the SM symmetry via radiative
corrections from renormalization group extrapolation of the scalar
Higgs masses from the GUT to TeV scale. As is well known, the
large top quark coupling enables us to achieve a similar goal i.e.
radiative EWSB in the case of MSSM~\cite{wise}. The simple
generalization of that procedure cannot work in our model since
the bidoublet Higgs of LR models contains both the $H_{u,d}$
components of MSSM, and as a result, large top quark coupling will
necessarily turn both their masses negative and this is known not
to give a stable vacuum.

Our proposal is that we use a domain of parameter space for the
soft SUSY-breaking mass squares for the RH Higgs doublets $\chi^c$
and $\bar{\chi}^c$ where the mass square of one of them turns
negative, by RG running to the TeV scale due to the $L^c\bar\chi^c
S$ Yukawa coupling being large. This leads to a breaking of the
$SU(2)_R$ and $B-L$ symmetry. The mass square of the $\chi^c$
remains positive throughout but it acquires an induced VEV.
The differences in their VEVs, via the $D$-term, can make the mass
square of the $H_u$ field negative while keeping the mass square
of $H_d$ positive as in the case of MSSM, thereby also giving rise
to the EWSB. The main point is that both symmetry breakings owe
their origin to one radiative correction.

In order to show that it is indeed possible to achieve negative mass square
for one of the RH Higgs doublets while keeping all other soft mass squares
positive, we need to examine the RG running of all the soft mass parameters
from the GUT to TeV scale. In this regime, the model is SUSYLR for which the
superpotential and soft SUSY-breaking Lagrangian are given by~\cite{nick}
\begin{eqnarray}
     W &=& ih_a Q^T \tau_2\Phi_a Q^c +
     ih'_a L^T \tau_2 \Phi_a L^c +
     i\mu^\alpha_{\chi^c_{pq}}S^\alpha \chi^{c^T}_p\tau_2 \bar{\chi}^c_q
     + i\mu^\alpha_{L^c_p}S^\alpha L^{c^T}\tau_2 \bar\chi^c_p\nonumber\\
     &&
     + iM_{\chi^c}\chi^{c^T}\tau_2 \bar{\chi}^c
     + \mu^\alpha_{\Phi_{ab}}S^\alpha{\rm Tr}
     \left(\Phi^T_a\tau_2\Phi_b\tau_2\right)
     + M_{\Phi_{ab}}{\rm Tr}\left(\Phi^T_a\tau_2\Phi_b\tau_2\right)
     \nonumber\\
     &&
     +\frac{1}{6}Y^{\alpha\beta\gamma}S^\alpha S^\beta S^\gamma
     + \frac{1}{2}M^{\alpha\beta}_S S^\alpha S^\beta,
     \label{eq:superpotential}\\
     {\cal L}_{\rm soft} &=&
     -\frac{1}{2}\left(M_3\tilde{g}\tilde{g}+M_{2L}\tilde{W}_L\tilde{W}_L
     +M_{2R}\tilde{W}_R\tilde{W}_R+M_1\tilde{B}\tilde{B}+{\rm h.c.}\right)
     \nonumber\\
     &&
     - \left[iA_{Q_a}\tilde{Q}^T\tau_2\Phi_a\tilde{Q}^c+
     iA_{L_a}\tilde{L}^T\tau_2\Phi_a\tilde{L}^c+iA_{\chi^c_{pq}}^\alpha
     S^\alpha \chi^{c^T}_p\tau_2\bar{\chi}^c_q+iA_{L^c_{p}}^\alpha
     S^\alpha \tilde{L}^{c^T}\tau_2\bar{\chi}_p^c\right.\nonumber\\
     && \left.
     +\frac{1}{6}A_S^{\alpha\beta\gamma}S^\alpha S^\beta S^\gamma +
     A^\alpha_{\Phi_{ab}}S^\alpha{\rm Tr}\left(\Phi^T_a\tau_2\Phi_b\tau_2
     \right)+{\rm h.c.}\right]\nonumber\\
     &&
     -\left[iB_{\chi^c_{pq}}\chi^{c^T}_p\tau_2\bar{\chi}^c_q
     + B_{ab}{\rm Tr}\left(\Phi^T_a\tau_2\Phi_b\tau_2\right)
     +\frac{1}{2}B_S^{\alpha\beta}S^\alpha S^\beta\right]\nonumber\\
     &&
     -\left[m_Q^2\tilde{Q}^T\tilde{Q}^*+
     m_{Q^c}^2\tilde{Q}^{c^\dag}\tilde{Q}^c+
     m_{L}^2\tilde{L}^T\tilde{L}^*
     +m_{L^c}^2\tilde{L}^{c^\dag}\tilde{L}^c
     +m_{\chi^c}^2\chi^{c^\dag}_p\chi^c_p+m_{\bar{\chi}^c}^2\bar{
     \chi}^{c^\dag}_p\bar{\chi}^c_p\right. \nonumber\\
     && \left. +
     m_{\Phi_{ab}}^2{\rm Tr}\left(\Phi_a^\dag \Phi_b\right)
     + m_{S_{\alpha\beta}}^2 S^{\alpha^*}S^\beta\right]
     \label{eq:softl}
 \end{eqnarray}
where we have suppressed the generational and $SU(2)$ indices, and
$a,b=1,2$ (for two bidoublets), $p,q=1,2$ (for two $SU(2)_R$
doublets) and $\alpha,\beta,\gamma=1,2,3$ (for three gauge
singlets). Note that we do not have any $\chi$-term in these
expressions as there is no $SU(2)_L$ Higgs doublet in our model.
Also we have an additional term in the superpotential (the
$SL^c\chi^c$ term) and a corresponding trilinear term in the soft
breaking Lagrangian (the $S\tilde{L}^c\bar{\chi}^c$ term) as compared to
the expressions given in Ref.~\cite{nick}; this additional term in
the superpotential is required for the inverse seesaw mechanism to
work. Moreover, if we assume $R$-parity conservation, then the
$S\chi^c\bar{\chi}^c$ and $S\Phi\Phi$ terms are not allowed in the
superpotential and also in the soft-breaking Lagrangian, i.e.
the couplings $\mu_{\chi^c}$ and $\mu_{\Phi}$
as well as $Y_{abc}$ in Eq.~(\ref{eq:superpotential}) and the corresponding
terms in Eq.~(\ref{eq:softl}) are set to
zero and $\mu_{L^c}$ is the only non-zero coupling in
Eq.~(\ref{eq:superpotential}) which can be fixed by requiring $b-\tau$
unification at the GUT-scale. In this
section, we work with this assumption; the effects of $R$- parity
breaking will be discussed later.

\begin{figure}[h!]
    \centering
    \includegraphics[width=10cm]{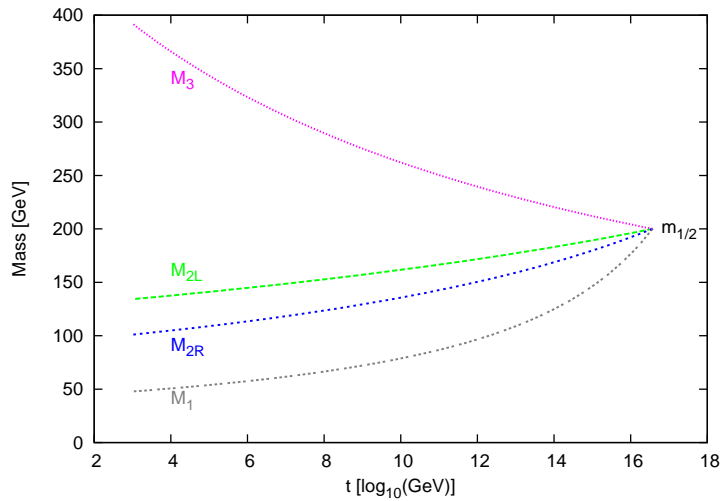}
    \caption{RG evolution of gaugino masses from GUT to TeV scale for
    $m_{1/2}=200$ GeV.}
    \label{fig:gaugino}
\end{figure}

Now we analyze the RG evolution of the gaugino and soft mass parameters from
GUT to TeV scale. It is well known that in SUSY GUTs,
the $\beta$-function for the gaugino mass is proportional
to the $\beta$-function for the corresponding gauge coupling. Explicitly, the
RGEs for the gaugino mass parameters are given by
\begin{eqnarray}
    \frac{dM_i}{dt} = \frac{2b_i}{16\pi^2}M_i g_i^2
    \label{eq:gaugino}
\end{eqnarray}
where the $\beta$-function coefficients in our SUSYLR model are~\cite{model}
$b_i = (13,2,4,-2)$, corresponding to $i=\mathbf 1_{B-L},~\mathbf 2_L,~
\mathbf 2_R,~\mathbf 3_c$ respectively. This implies that the three gaugino
masses, like the three gauge couplings, must unify at $\mu=M_{\rm GUT}$. In
order to solve Eq.~(\ref{eq:gaugino}), we adopt the universality hypothesis at
the GUT scale (as in typical mSUGRA type models)
\begin{eqnarray}
    M_1 = M_{2L} = M_{2R} = M_3 \equiv m_{1/2},
\end{eqnarray}
together with the initial condition
\begin{eqnarray}
    g^2_1=g^2_{2L}=g^2_{2R}=g^2_3\equiv 4\pi\alpha_{\rm GUT},
\end{eqnarray}
where $M_{\rm GUT}\simeq 4\times 10^{16}~{\rm GeV}$ and
$\alpha_{\rm GUT}^{-1}\simeq 20.3$ in our model~\cite{model}. Using these
initial conditions, we can obtain the running masses for the gauginos at TeV
scale, starting with a given value $m_{1/2}$ at the GUT scale, as shown in
Fig.~\ref{fig:gaugino} for a typical value of $m_{1/2}=200$ GeV. The value of
$M_3$ increases, since it has a negative $\beta$-function, while the other
gaugino masses decrease as we go down the energy scale. Thus the gluino is
much heavier than other gauginos at the weak scale.

The one-loop RGEs for the soft SUSY-breaking mass parameters are given in
Appendix A. As initial conditions, we assume universality and reality of the
soft fermion and Higgs masses at the GUT-scale, i.e.
\begin{eqnarray}
    &&\left(m^2_Q\right)_{ij} = \left(m^2_{Q^c}\right)_{ij} =
    \left(m^2_{L}\right)_{ij} = \left(m^2_{L^c}\right)_{ij}
    \equiv m^2_0 \delta_{ij},\nonumber\\
    && m^2_{\chi^c} = m^2_{\bar{\chi}^c} = m^2_0,~
    \left(m^2_\Phi\right)_{ab} = m^2_0 \delta_{ab},
    \label{eq:softin}
\end{eqnarray}
whereas a different scale is assumed for the soft singlet scalar mass:
\begin{eqnarray}
    \left(m^2_S\right)_{\alpha\beta} = m'^{2}_0~ ~ ~ ~\forall~
    \alpha,\beta = 1,2,3.
\end{eqnarray}
In principle, we can choose a different mass scale for the Higgs
bidoublets and even different generations of fermions as well. The
only constraint due to the $SO(10)$ symmetry requires us to have
the same mass for each generation of fermions. Note that all the
off-diagonal soft SUSY breaking scalar masses have been set to
zero. The inter-generation mixing at the low energy scale then
occurs only via the superpotential Yukawa couplings. With these
initial conditions, we solve the coupled RGEs for the soft masses
given in Appendix A, along with the Yukawa RGEs given in Ref.~\cite{model},
to get the
running soft masses at the low scale. We find that it is indeed
possible to find a parameter space such that $m^2_{\bar\chi^c}<0$
(for $SU(2)_R$ breaking) and $m^2_{\Phi_1} < 0$ (for EWSB) while keeping
all other mass squares positive.
\begin{figure}[h!]
    \centering
    \includegraphics[width=10cm]{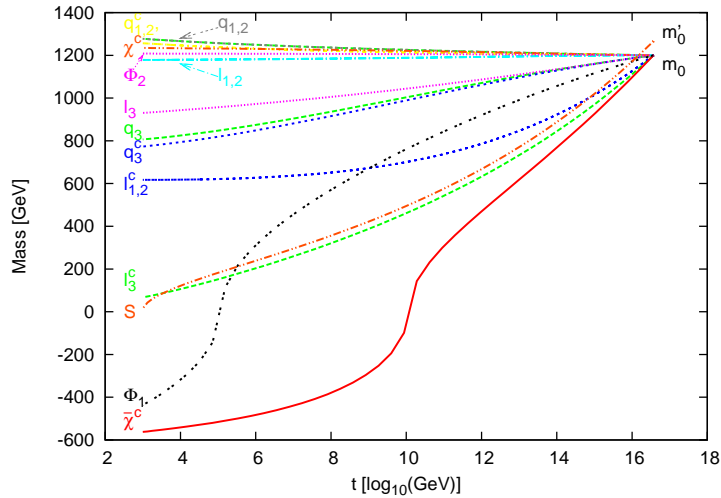}
    \caption{Evolution of the scalar mass parameters for $m_{1/2} = 200$ GeV, 
    $m_0 = 1.20$ TeV and $m'_0 = 1.27$ TeV. 
    For the scalar masses, we actually plot
    sign$(m^2)\cdot\sqrt{|m^2|}$, so that the negative values on the curves
    correspond to negative values of $m^2$.}
    \label{fig:soft}
\end{figure}
Fig.~\ref{fig:soft} illustrates such a scenario for the choice
$m_{1/2}=200$ GeV, $m_0 = 1.2$ TeV and $m'_0 = 1.27$ TeV. We have chosen the 
$SL^c\bar\chi^c$ coupling $\mu_{L^c}=0.7$ to achieve a realistic fermion mass 
spectrum, and in particular, the $b-\tau$ unification at the GUT scale.  Note
that the RH slepton masses evolve much more rapidly than their LH
counterparts due to this large coupling $\mu_{L^c}$. The value of
$m'_0$ is chosen such that all the other
eigenvalues (especially $m_{L_3^c}^2$ and $m^2_{S}$) remain
positive at the TeV scale. Note that the low energy values of $m_{L_3^c}^2$
and $m^2_{S}$ are of order (10 GeV)$^2$. However the physical masses of
these particles also receive a contribution from the
$\langle\bar\chi^c\rangle$ which pushes the masses upto a TeV scale. As far as
the squark masses are concerned, they evolve more than the slepton
masses due to the strong interaction loop contributions to their
RGEs. The small intra-generational mass splitting
    is due to the differences in their electroweak interaction. We can see
    clearly that at the weak scale, the values of $m^2_{\bar \chi^c}$ and
    $m^2_{\Phi_1}$ are negative, thus triggering the $SU(2)_R$ and electroweak
    symmetry breaking respectively. Note that we need not have both the
    bidoublet mass squares to be negative, as one negative value will induce
    the symmetry breaking via the cross terms of the type $\Phi_1\Phi_2$ in the
    Lagrangian.

We also verify that the low-energy values of the sfermion mass square matrices
 satisfy all the FCNC constraints~\cite{masiero}, due to the smallness of
the off-diagonal entries. As an example, we give the values here for the
parameter values shown in Fig. 2:
{\small \begin{eqnarray}
    m_Q^2 &=& \left(\begin{array}{ccc}
    1.63\times 10^6 & -1.45\times 10^1+8.64\times 10^1i &
    -4.79\times 10^2+3.57\times 10^3i\\
    -1.45\times 10^1-8.64\times 10^1i & 1.63\times 10^6 &
    -2.31\times 10^4+1.68i\\
    -4.79\times 10^2-3.57\times 10^3i & -2.31\times 10^4-1.68i &
    6.51\times 10^5
\end{array}\right)~{\rm GeV}^2,\nonumber\\
m^2_{Q^c} &=& \left(\begin{array}{ccc}
    1.58\times 10^6 & -1.45\times 10^1+8.64\times 10^1i & -4.79\times 10^2+3.57
    \times 10^3i\\
    -1.45\times 10^1-8.64\times 10^1i & 1.58\times 10^6 &
    -2.31\times 10^4+1.68i\\
    -4.79\times 10^2-3.57\times 10^3i & -2.31\times 10^4-1.68i &
    5.99\times 10^5
\end{array}\right)~{\rm GeV}^2,\nonumber\\
m^2_L &=& \left(\begin{array}{ccc}
1.39\times 10^6 & -7.28+8.39\times 10^1i & -2.59\times 10^2+3.45\times 10^3i\\
-7.28-8.39\times 10^1i & 1.39\times 10^6 &
-1.25\times 10^4+7.45\times 10^{-1}i\\
-2.59\times 10^2-3.45\times 10^3i & -1.25\times 10^4-7.45\times 10^{-1}i &
8.66\times 10^5
\end{array}\right)~{\rm GeV}^2,\nonumber\\
m^2_{L^c} &=& \left(\begin{array}{ccc}
    3.81\times 10^5 & -7.18+8.24\times 10^1i &
    -2.57\times 10^2+3.41\times 10^3i \\
    -7.18-8.24\times 10^1i & 3.81\times 10^5 &
    -1.24\times 10^4+7.75\times 10^{-1}i \\
    -2.57\times 10^2-3.42\times 10^3i & -1.24\times 10^4-7.75\times 10^{-1}i &
    5.00\times 10^3
\end{array}\right)~{\rm GeV}^2.\nonumber
\end{eqnarray}}
\section{Proton decay}
In this section, we discuss the partial lifetimes of various
proton decay channels.
\subsection{Proton decay operators}
\begin{figure}[h!]
    \centering
    \includegraphics[width=5cm]{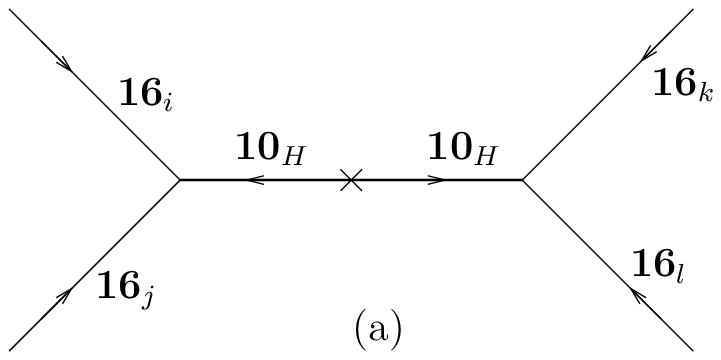}
    \hspace{4cm}
    \includegraphics[width=5cm]{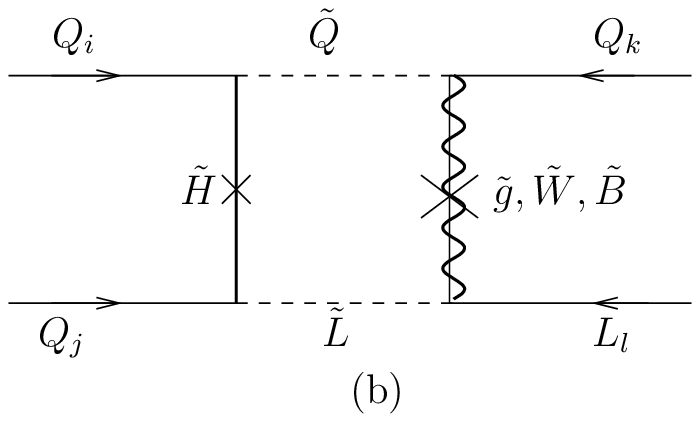}
    \caption{(a) Supergraph giving rise to effective dimension-5 proton
    decay operators, and (b) Box diagram involving gaugino exchange that
    converts the dimension-5 operator of Fig. 3(a) into an effective four-
    Fermi operator that induces proton decay.}
    \label{fig:dim5}
\end{figure}

In generic SUSY-GUTs, there exist three sources for proton decay:
\begin{itemize}
    \item $D$-type (dimension-6) operators that arise from exchange
of gauge bosons:
        \begin{equation}
            \frac{1}{M_G^2}\int d^2\theta~d^2\overline{\theta}~
            \Phi^\dagger \Phi \Phi^\dagger \Phi,
        \end{equation}
        which may be generated both by heavy gauge boson exchange and
        by heavy chiral (Higgs) superfield exchange. For a unification
        scale $\gsim 10^{16}$ GeV, these contributions to proton decay
        are sufficiently small and well beyond the range of current
        experiments.

    \item $F$-type (dimension-5) operators that arise from the
exchange of color triplet Higgsino fields in {\bf 10}-Higgs fields as
shown in Fig. \ref{fig:dim5}(a):
        \begin{equation}
            \frac{1}{M_G}\int d^2\theta~\Phi\Phi\Phi\Phi
            \label{eq:dim5}
        \end{equation}
        where $\Phi$'s are used to denote quark and lepton
        doublets.
         In the component language, they give rise
        to dimension-5 operators of the form
        $(QQ)(\tilde{Q}\tilde{L})$ and $(QL)(\tilde{Q}\tilde{Q})$. As
        these operators involve squark and slepton fields, they cannot
        induce proton decay in the lowest-order. Proton decay occurs
        by converting the squark and slepton legs into quarks and
        leptons by exchanging a gaugino, as shown in the box
    diagram of Fig. \ref{fig:dim5}(b).
          \item Another class of dimension-5 operators arising from
$R$-parity breaking Planck suppressed operators, which are absent when we
assume $R$-parity. We discuss them in Sec. VI and show that their effects
are very small due to low $B-L$ breaking scale. These are absent
in models where {\bf 126} Higgs fields break $B-L$, but are present in our
model.
\end{itemize}

There are two effective dimension-5 operators of $LLLL$ type that
involve only left-handed quark and lepton fields, given by
Eq.~(\ref{eq:dim5}) and a corresponding $RRRR$ type, both
invariant under MSSM~\cite{dim5}. In super-space notation, these
are explicitly given by
\begin{eqnarray}
    {\cal O}_{L} &=& \int d^2\theta~\epsilon^{\alpha\beta\gamma}
    \epsilon^{ab}\epsilon^{cd}~Q_{\alpha ai}Q_{\beta bj}
    Q_{\gamma ck}L_{dl}~,
    \label{eq:ol}\\
    {\cal O}_R &=& \int d^2{\theta}~\epsilon^{\alpha\beta\gamma}
    \left(Q^c\right)_{\alpha i}
    \left(Q^c\right) _{\beta j}
    \left(Q^c\right) _{\gamma k}
    \left(L^c\right) _{l}
    \label{eq:or}
\end{eqnarray}
where $\alpha,\beta,\gamma=1,2,3$ are $SU(3)_c$ color
indices;
$a,b,c,d=1,2$ are $SU(2)_L$ isospin indices; and $i,j,k,l=1,2,3$ are
generation indices. It is clear from the form of these operators that they
break baryon number by one unit, but preserve the $B-L$ symmetry, leading to
the proton decay to a pseudoscalar and an anti-lepton.
As argued in Ref.~\cite{ibanez} for kinematical reasons and explicitly shown
in Ref.~\cite{goto} for small to moderate $\tan\beta$ region of the SUSY
parameter space, the $RRRR$ contributions are at least an order of magnitude
smaller than the $LLLL$ contributions. We also verify this in our model, as
shown later; for the time being therefore, we concentrate only on the $LLLL$
operator.

\begin{figure}[h!]
    \centering
    \includegraphics[width=5cm]{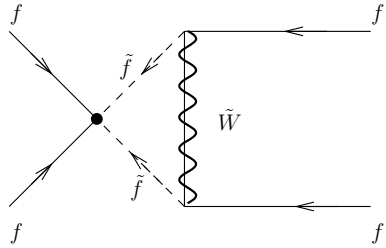}
    \caption{The effective four-Fermi interaction diagram induced by the
    gaugino dressing of the effective dimension-5 operator given by Fig.
    \ref{fig:dim5}(b).}
    \label{fig:4fermi}
\end{figure}

In component form, the effective superpotential due to the $LLLL$ operator is
explicitly given by~\cite{goh}
\begin{eqnarray}
    {\cal W}_{\Delta B=1} = \frac{1}{M_T}\epsilon^{\alpha\beta\gamma}
    \left[\left(C_{ijkl}-C_{kjil}\right)u_{\alpha i}d_{\beta j}u_{\gamma k}
    e_l - \left(C_{ijkl}-C_{ikjl}\right)u_{\alpha i}d_{\beta j}d_{\gamma k}
    \nu_{l}\right]
    \label{eq:l}
\end{eqnarray}
where $M_T$ is the effective mass of the color triplet Higgs field belonging
to the ${\bf 10}_H$ representation, and in our model, is of the order of the
unification scale $M_G$ (see Appendix A of Ref.~\cite{model}). This
superpotential leads to the effective dimension-5 operators involving two
fermions and two sfermions as shown in Fig.~\ref{fig:dim5}(b),
which lead to proton decay by four-Fermi
interactions when ``dressed'' via the exchange of gauginos, namely gluinos,
binos and winos. A typical diagram
for the effective four-Fermi interaction induced by this dressing is shown in
Fig.~\ref{fig:4fermi}.

The coefficients $C_{ijkl}$ associated with the superpotential given by
Eq.~(\ref{eq:l}) can be expressed in terms of the products of the GUT-scale
Yukawa couplings. For model (A), this is given by
\begin{eqnarray}
    C_{ijkl} &=& h_{u_{ij}}h_{u_{kl}}
    +x_1h_{d_{ij}}h_{d_{kl}}+x_2h_{u_{ij}}h_{d_{kl}}+x_3h_{d_{ij}}h_{u_{kl}}
    +\frac{1}{2}\left[h_{u_{ij}}f_{u_{kl}}+f_{u_{ij}}h_{u_{kl}}\right.
    \nonumber\\
    &&
    \left.+x_1\left(h_{d_{ij}}f_{d_{kl}}+f_{d_{ij}}h_{d_{kl}}\right)
    +x_2\left(f_{u_{ij}}h_{d_{kl}}+h_{u_{ij}}f_{d_{kl}}\right)
    +x_3\left(h_{d_{ij}}f_{u_{kl}}+f_{d_{ij}}h_{u_{kl}}\right)\right]
    \nonumber\\
    &&
    +\frac{1}{4}\left(f_{u_{ij}}f_{u_{kl}}+x_1f_{d_{ij}}f_{d_{kl}}+
    x_2f_{u_{ij}}f_{d_{kl}}+x_3f_{d_{ij}}f_{u_{kl}}\right)
    \label{eq:coeffa}
\end{eqnarray}
while for model (B) this becomes
\begin{eqnarray}
    C_{ijkl} &=& h_{ij}h_{kl}+x_1h'_{ij}h'_{kl}+x_2h_{ij}h'_{kl}
    +x_3h'_{ij}h_{kl}+\frac{1}{2}\left[x_1\left(h'_{ij}f_{kl}
    +f_{ij}h'_{kl}\right)+x_2h_{ij}f_{kl}+x_3f_{ij}h_{kl}\right]\nonumber\\
    &&
    +\frac{1}{4}x_1f_{ij}f_{kl}
    \label{eq:coeffb}
\end{eqnarray}
where $x_i$'s are the ratios of the $\mathbf {10}_H$ color triplet Higgs
masses and mixings and the factor $\frac{1}{2}$ is the C-G coefficient for the
$\mathbf {10\cdot 10\cdot 126}$ coupling. Note that there are only three mixing
parameters as there are only four color triplet Higgses in the
MSSM gauge group, corresponding to the two ${\bf 10}_H$ fields in our model.
As we are interested only in the upper bound for the partial lifetimes of
various proton decay channels, we do not need to know the detailed form for
the $x_i$ parameters in terms of these masses and mixings. We just vary
these parameters numerically to get the maximum value for the partial lifetimes.

It can be shown that~\cite{gluino} in the limit of all squark masses
being
degenerate as in typical mSUGRA type models, the gluino and bino contributions
to the dressing of the dimension-5 operators vanish. This basically follows
from the use of Fierz identity for the chiral two component spinors
representing quarks and leptons. In realistic models,
the FCNC constraints allow only very small
deviations from universality of squark masses. Hence, these gluino and bino
contributions are expected to be small compared to the wino
contributions, and can be ignored altogether. The charged wino dressing
diagrams have
been evaluated earlier~\cite{nath}, and in the limit of degenerate squark
masses, this leads to the effective Lagrangian~\cite{goh}
\begin{eqnarray}
    {\cal L}_{\Delta B=1}=2I\epsilon^{\alpha\beta\gamma}(C_{kjil}-C_{ijkl})
    [u_{\alpha k}^T C d_{\beta j} d_{\gamma i}^T C \nu_l+
    u_{\beta j}^T C d_{\gamma k} u_{\alpha i}^T C e_l],
    \label{eq:lag}
\end{eqnarray}
where $C$ denotes the charge-conjugation matrix and $I$ is given by
\begin{equation}
    I = \frac{\alpha_2}{4\pi}\frac{m_{\tilde{W}}}{M^2_{\tilde{f}}},
\end{equation}
$m_{\tilde{W}}$ being the wino mass and $M_{\tilde{f}}$ the
sfermion mass. Using this expression and adding a similar
contribution from the neutral wino exchange diagram, we can write
down the total contribution to various proton decay channels. This
is summarized in Table \ref{tab:coeff}. We note that the proton
decay operators with $s$-quark lead to $K$-meson final states
whereas the ones without $s$ lead to $\pi$ final states. As shown
in Table \ref{tab:coeff}, the amplitude for non-strange quark final
states will be Cabibbo-suppressed compared to the strange quark final
states. It is also important to mention here that the total
amplitude for final states involving neutrinos is the incoherent
sum of the rates for all three neutrino states. This leads to
large decay rates for $p\to K^+\overline{\nu}$ and $p\to
\pi^+\overline{\nu}$ channels compared to the other decay channels
due to the large Yukawa couplings of the third generation.
\begin{table}[h!]
\begin{center}
    \begin{tabular}{||c|c||} \hline\hline
        Decay channel & ${\cal C}$-coefficient \\
        \hline\hline
        $p\to K^+\overline{\nu}_l$ &
        $(C_{112l}-C_{121l})$ \\
        $p\to K^0e^+$ &
        $(C_{1121}-C_{1211})$\\
        $p\to K^0\mu^+$ &
        $(C_{1122}-C_{1212})$\\
        $p\to \pi^+\overline{\nu}_l$ &
        $\sin{\theta_C}(C_{211l}-C_{112l})$ \\
        $p\to \pi^0e^+$ &
        $\sin{\theta_C}(C_{2111}-C_{1121})$ \\
        $p\to \pi^0\mu^+$ &
        $\sin{\theta_C}(C_{2112}-C_{1122})$\\
        \hline\hline
    \end{tabular}
\end{center}
\caption{The coefficients for various $\Delta B=1$ dimension-5 operators
obtained from the effective Lagrangian to leading order. Here $\theta_C$ is the
Cabibbo angle (with $\sin\theta_C\sim 0.22$) and the $C_{ijkl}$'s are
products of the Yukawa couplings, as defined in Eqs.~(\ref{eq:coeffa}) and
(\ref{eq:coeffb}).}
\label{tab:coeff}
\end{table}

Before proceeding to calculate the rate of proton decay induced by
these $LLLL$ type operators, let us estimate the contribution from
the $RRRR$ type operators in our model. The gluino dressing graphs
do not contribute in the limit of universal sfermion masses by the
same Fierz arguments as for the $LLLL$ case. Moreover, since all
superfields in the $RRRR$ operator are $SU(2)_L$ singlets, there
is no wino contribution to the leading order. Also the bino
dressing generates an effective four-Fermi operator of the type
$\epsilon^{\alpha\beta\gamma}\epsilon^{ij}\epsilon^{kl}
u^{c^T}_{\beta j}C d^c_{\gamma k}u^{c^T}_{\alpha i}Ce^c_l$ which,
in flavor basis, is antisymmetric in the flavor indices $i$ and
$j$, and hence in the mass basis, must involve a charm quark. Thus
to leading order, the bino contribution also vanishes due to phase
space constraints. Thus the only dominant contribution comes from
the Higgsino exchange and the largest amplitude in this case,
which comes from stop intermediate states, is estimated to
be~\cite{goh} (using the $C_{ijkl}$ values calculated later in our model)
\begin{eqnarray}
    C_{1323}\frac{m_t m_\tau V_{ub}}{16\pi^2 v_{\rm wk}^2\sin\beta\cos\beta} \sim
    4.0\times 10^{-10}
\end{eqnarray}
for $\tan\beta=30$, as compared to the $LLLL$ contribution which is typically of order
\begin{eqnarray}
    C_{1123}\frac{\alpha_2}{4\pi} \sim 4.5\times 10^{-9}
\end{eqnarray}
As the $RRRR$ contribution is proportional
to $\frac{1}{\sin\beta\cos\beta}$ which is $\sim\tan\beta$ for large $\beta$, for
smaller $\tan\beta$, this contribution is further suppressed.
This justifies why we can ignore the $RRRR$ contributions in the
following calculation of proton decay rate.
\subsection{Proton decay rate}
In order to calculate the proton decay rate, we must extrapolate these
dimension-5 operators defined at the GUT scale to the scale of $m_p=1$ GeV.
In our model, we can divide this whole energy range into three parts,
following the breaking chain given by Eq.~(\ref{eq:chain}):\\
(a)
from the GUT scale $M_G$ to the $B-L$ breaking scale $M_R$ (SUSYLR),\\
(b) from $M_R$
to the SUSY-breaking scale $M_S$ (MSSM), and \\
(c) from $M_S$ to 1 GeV (SM). \\
The values of these extrapolation factors are given in the literature~
\cite{ibanez,buras,su5a,su5b} for both SM
and MSSM, but not for the SUSYLR model. In this section, we derive these
factors using the anomalous dimensions for the dimension-5 operators in our
model calculated in Appendix B. We denote the overall extrapolation factor by
$A_e$. We noted some discrepancies in the values of the
anomalous dimensions quoted in different papers, but found that our
results for the SM and MSSM cases agree with those given in
Refs.~\cite{buras,ibanez} and quoted in Appendix E of Ref.~\cite{review}.

We also need to include the QCD effects in going from three quarks to proton.
As the low-energy hadrons are involved in the decay, this is a highly non-
perturbative process, and it is difficult to calculate the exact form of the
hadronic mixing matrix element for the process. Even though various QCD models
have been constructed for the purpose, the estimates vary by a factor of
${\cal O}(10)$ between the smallest and the largest~\cite{qcd1}. As the partial width
of the decay is proportional to the matrix element squared, the variation
in the estimate of proton lifetime in different models will be ${\cal O}(100)$.
A different approach using lattice QCD techniques gives more consistent
results~\cite{lattice}. We use these recent results to estimate the chiral
symmetry
breaking effects which can be parametrized by two hadronic parameters
$D$ and $F$. Then the hadronic mixing matrix for the proton decay can be
written as $\frac{\beta}{f_\pi}f(F,D)$
 where $f_\pi= (130.4 \pm 0.04 \pm 0.2) $ MeV~\cite{pdg} is the pion decay
constant and $|\beta|=0.0120(26)~{\rm GeV}^3$~\cite{lattice}
is a low-energy parameter of the $SU(3)_f$ baryon chiral
Lagrangian with the baryon number violating interaction. The factors $f(F,D)$
for different final states are listed in Appendix C.

Finally, combining all the factors discussed above, the proton decay rate for
a given decay mode $p\to Ml$ ($M$ denotes the meson and $l$ the lepton) is
given by~\cite{goh}
\begin{eqnarray}
    \Gamma_p (Ml) &\simeq& \frac{m_p}{32\pi M_T^2}\frac{|\beta|^2}{f_\pi^2}
    \left(\frac{\alpha_2}{4\pi}\right)^2
    \left(\frac{m_{\tilde{W}}}{M^2_{\tilde{f}}}\right)^2 4|{\cal C}|^2
    |A_e|^2 |f(F,D)|^2
    \nonumber\\
    &\simeq& \left(1.6\times 10^{-49}~{\rm GeV}\right)
    \left(\frac{2\times 10^{16}~{\rm GeV}}{M_T}\right)^2
    \left(\frac{m_{\tilde{W}}}{200~{\rm GeV}}\right)^2
    \left(\frac{1~{\rm TeV}}{M_{\tilde{f}}}\right)^4\nonumber\\
    && \times ~ |{\cal C}|^2|A_e|^2|f(F,D)|^2
    \label{eq:pml}
\end{eqnarray}
where the coefficients ${\cal C}$ are given in Table~\ref{tab:coeff}, the hadronic factors
$f(F,D)$ are listed in Appendix C, and the
extrapolation factors $A_e$ are derived below.
\subsection{The extrapolation factors for the dimension-5 operator}
As noted in the previous section, we need to extrapolate the dimension-5
operators defined at the GUT scale to the scale of 1 GeV. In our model, this
whole energy range is divided into three parts, with different running
behavior for the gauge couplings. First, we have the SM sector from
1 GeV to the SUSY-breaking scale $M_S$ in which we have the usual non-SUSY
enhancement factor~\cite{buras} for the $LLLL$ operator:
\begin{eqnarray}
    A^{\rm NS}_e = \left[\frac{\alpha_3(1~{\rm GeV})}{\alpha_3(M_S)}\right]
    ^{2/\left(11-\frac{2}{3}n_f\right)}
\end{eqnarray}
where $n_f$ is the number of quark flavors below the energy scale of interest.
Here we have neglected the effects of $SU(2)_L$ and $U(1)_Y$ couplings as they
are much smaller compared to that of $SU(3)_c$. In our model,
as $M_S=300~{\rm GeV}>m_t$, the enhancement factor explicitly becomes
\begin{eqnarray}
    A^{\rm NS}_e &=&
    \left[\frac{\alpha_3(1~{\rm GeV})}{\alpha_3(m_c)}\right]
    ^{2/9}
    \left[\frac{\alpha_3(m_c)}{\alpha_3(m_b)}\right]
    ^{6/25}
    \left[\frac{\alpha_3(m_b)}{\alpha_3(m_t)}\right]
    ^{6/23}
    \left[\frac{\alpha_3(m_t)}{\alpha_3(M_S)}\right]
    ^{2/7}= 1.49
\end{eqnarray}
using the values of $\alpha_3(\mu)$ at $\mu=1$ GeV, $m_c$ and $m_b$ obtained by
interpolating the renormalization group equation for the effective
QCD coupling~\cite{lbl} and at $\mu=m_t$ by the SM running from $\mu=m_Z$.

Now above $M_S$, we have the usual MSSM till the $B-L$ breaking scale $M_R$
and then the SUSYLR model till the GUT scale $M_G$. The extrapolation factor
in this case is given by
\begin{equation}
    A^{\rm S}_e = A^{\rm MSSM}_e A^{\rm SUSYLR}_e
\end{equation}
where the corresponding factors in the two sectors are given by
\begin{eqnarray}
    A^{\rm MSSM}_e = \prod_{i=1}^3
    \left[\frac{\alpha_i(M_S)}{\alpha_i(M_R)}\right]^
    {\frac{\gamma_i}{b_i}},~ ~{\rm and}~ ~
    A^{\rm SUSYLR}_e = \prod_{j=1}^4
    \left[\frac{\alpha_j(M_R)}{\alpha_j(M_G)}\right]^
    {\frac{\gamma_j}{b_j}}
\end{eqnarray}
Here $b_i=\left(\frac{33}{5},1,-3\right)$ for $i=\mathbf 1_Y,\mathbf 2_L,\mathbf 3_c$ are
the well known MSSM $\beta$-function coefficients,
$b_j=(13, 2, 4, -2)$ for $j=\mathbf 1_{B-L},\mathbf 2_L,\mathbf 2_R,\mathbf 3_c$ are
the $\beta$-function coefficients for the SUSYLR model~\cite{model}, and
$\gamma_i$'s are the anomalous dimensions for the $LLLL$ operator, calculated
in Appendix B. Using these results, we obtain
\begin{eqnarray}
    A^{\rm MSSM}_e &=&
    \left[\frac{\alpha_3(M_S)}{\alpha_3(M_R)}\right]^{-4/3}
    \left[\frac{\alpha_{2_L}(M_S)}{\alpha_{2_L}(M_R)}\right]^{3}
    \left[\frac{\alpha_{1_Y}(M_S)}{\alpha_{1_Y}(M_R)}\right]^{1/33}
    = 0.91
\end{eqnarray}
using the MSSM running of the gauge couplings, and similarly,
\begin{eqnarray}
    A^{\rm SUSYLR}_e &=&
    \left[\frac{\alpha_3(M_R)}{\alpha_3(M_G)}\right]^{-2}
    \left[\frac{\alpha_{2_L}(M_R)}{\alpha_{2_L}(M_G)}\right]^{3/2}
    \left[\frac{\alpha_{2_R}(M_R)}{\alpha_{2_R}(M_G)}\right]^{3/4}
    \left[\frac{\alpha_{1_{B-L}}(M_R)}{\alpha_{1_{B-L}}(M_G)}\right]^{1/26}
    = 0.08
\end{eqnarray}
using the SUSYLR running of the gauge couplings~\cite{model}.
Combining all these results, we get the overall extrapolation factor in
bringing the operators from the GUT scale down to 1 GeV:
\begin{eqnarray}
    A_e = A^{\rm NS}_e A^{\rm MSSM}_e A^{\rm SUSYLR}_e = 0.11
    \label{eq:Ae}
\end{eqnarray}
\subsection{Predictions for partial lifetimes}
Substituting the extrapolation factor obtained in Eq.~(\ref{eq:Ae}) in the
expression for the partial decay width given by Eq.~(\ref{eq:pml}) and using
$M_T\simeq M_U\simeq 4\times 10^{16}~{\rm GeV}$ in our model, we obtain
the partial lifetimes of different decay modes:
\begin{eqnarray}
    \tau_p (Ml) = \frac{\hbar}{\Gamma_p} \simeq
    \frac{\left(4.42\times 10^{33}~{\rm years}\right)}
    {|f(F,D)|^2}\left(\frac{10^{-14}}{
    |{\cal C}|^2}\right)
    \left(\frac{200~{\rm GeV}}{m_{\tilde{W}}}\right)^2
    \left(\frac{M_{\tilde f}}{1~{\rm TeV}}\right)^4
    \label{eq:tpml}
\end{eqnarray}
The wino mass, $m_{\tilde{W}}$, has been constrained at LEP to be larger
than $\sim$ 100 GeV~\cite{grivaz},
essentially independent of any specific model.
As a typical value, we choose the universal gaugino mass, $m_{1/2}=200$ GeV,
which
when extrapolated to the weak scale gives $m_{\tilde{W}}\simeq 134$ GeV for
the wino mass.
\subsection*{Model (A)}
\begin{figure}[h!]
    \centering
    \includegraphics[width=10cm]{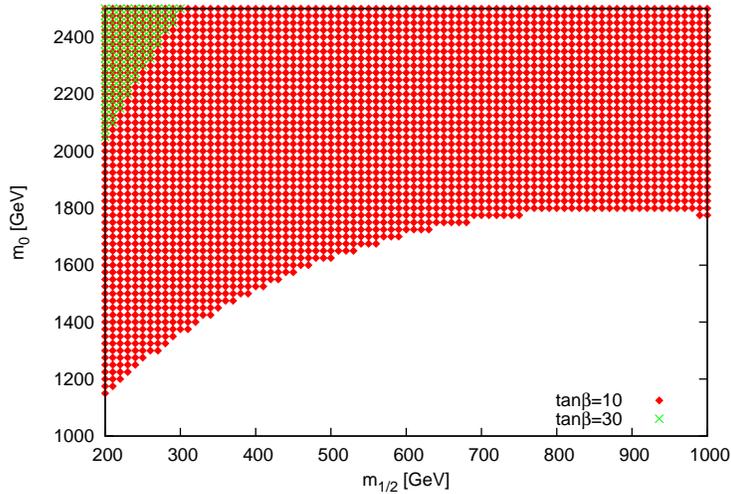}
    \caption{Model (A) allowed region in the $m_0-m_{1/2}$ plane satisfying the
    proton decay and EWSB constraints for $\tan\beta=10$ (red)
    and $\tan\beta=30$ (green).}
    \label{fig:moda}
\end{figure}

As we are interested in obtaining an upper
bound on the partial lifetimes of various proton decay modes, we adopt the
strategy of varying the mixing parameters $x_i$'s defined by Eq.~(\ref{eq:coeffa})
to maximize the expression~(\ref{eq:tpml}) and simultaneously satisfying the
present experimental lower bounds~\cite{kamio}. We find that the most
stringent constraint comes from the $p\to K^+\overline{\nu}$ decay mode, and
for this decay rate to be consistent with the present
experimental bound, we must have the sfermion mass $M_{\tilde{f}}\geq 1.2~
(2.1)$ TeV for the MSSM $\tan{\beta}=10~(30)$.
This value of $M_{\tilde{f}}$, when extrapolated to the GUT-scale,
 puts a lower limit on the universal squark mass $m_0$ for a given value of
$m_{1/2}$. The allowed region in the $m_0-m_{1/2}$ plane satisfying the proton
decay constraints and also satisfying the EWSB constraints is shown in Fig.
\ref{fig:moda}. It is clear that this model favors low values of $\tan\beta$.
\begin{table}[h!]
    \begin{center}
        \begin{tabular}{||c|c|c|c||}\hline\hline
        Decay & Experimental & \multicolumn{2}{c||}
        {Predicted upper limit ($\times 10^{33}$ yr)}\\ \cline{3-4}
        mode & lower limit ($\times 10^{33}$ yr) & $\tan{\beta}
        =10$ & $\tan{\beta}=30$
        \\
            \hline\hline
            $p\to K^+\overline{\nu}$ & 2.3 & 2.3 & 2.3\\
            $p\to K^0\mu^+$ & 1.3 & 399.3 & 738.8\\
            $p\to K^0 e^+$ & 1.0 & $1.3\times 10^3$ & 49.7\\
            $p\to \pi^0 e^+$ & 10.1 & $5.8\times 10^3$ & 230.0\\
            $p\to \pi^0 \mu^+$ & 6.6 & $2.4\times 10^4$ & $1.3\times 10^4$\\
            $p\to \pi^+ \overline{\nu}$ & 0.025 & 1.5 & 0.8\\
            \hline\hline
        \end{tabular}
    \end{center}
    \caption{Model (A) predictions for the upper limits on the partial
    lifetimes of various proton decay modes in $SO(10)$ with low scale SUSYLR
    for $\tan{\beta}=10$ and 30 for $m_{1/2}=200$ GeV. We have chosen
    the value of the universal scalar mass $m_0$ to be 1.2 (2.1) TeV for
    $\tan\beta=10~(30)$ so that the $p\to K^+
    \overline{\nu}$ constraint is just satisfied. The present experimental
    lower limits are also given for comparison.}
    \label{tab:moda}
\end{table}

The model predictions for the upper bound on partial lifetime of
various proton decay modes are given in Table \ref{tab:moda}. We
also list the present experimental lower bounds for comparison. As
noted above, the most stringent constraint on the parameter space
comes from the $p\to K^+\overline{\nu}$ decay mode; this is due to
the fact that the neutrino final states add incoherently for the
three generations, and hence, the decay rate for the neutrino
final states will be much larger compared to the rates of other
decay modes due to the third generation Yukawa coupling dominance.
This also explains why the $p\to \pi^+\overline{\nu}$ decay rate
is so large, even though it is Cabibbo-suppressed. The predicted
upper bounds for these neutrino final states may be testable in
the future proton decay searches, as in the next round of
Super-Kamiokande~\cite{kamio} or megaton type detector searches.
\subsection*{Model (B)}
As in the model (A), we maximize the function $|C|^{-2}$ given by
Eq.~(\ref{eq:coeffb}) with respect to the $x_i$ parameters to find
an upper bound on the proton decay lifetime. However, due to the
particular structure of the Yukawa matrices in this model, as
given by Eqs.~(\ref{eq:byuk10}) and (\ref{eq:byuk30}), the
parameters $x_2$ and $x_3$ have no effect on the amplitude and the
only effective mixing parameter is $x_1$. The experimental lower
bounds on the lifetime of various proton decay modes will then put
a lower bound on the ratio $\frac{M^2_{\tilde{f}}}
{x_1m_{\tilde{W}}}$. It turns out that the most stringent bound is
$p\to K^+\bar{\nu}~(\pi^0\mu^+)$ for $\tan\beta=10~(30)$
and we must have
\begin{eqnarray}
    \frac{M^2_{\tilde{f}}}{x_1m_{\tilde{W}}}\geq
    1.44~ (1.06)\times 10^5~{\rm GeV}
    \end{eqnarray}
As an example, for $m_{1/2}=200$ GeV and $x_1=0.1$, it puts a lower bound on the
first and second generation squark masses to be $M_{\tilde{f}}\geq 1.4~(1.2)$
TeV for $\tan\beta=10~(30)$.
\begin{table}[h!]
    \begin{center}
        \begin{tabular}{||c|c|c|c||}\hline\hline
        Decay & Experimental & \multicolumn{2}{c||}
        {Predicted upper limit ($\times 10^{33}$ yr)}\\ \cline{3-4}
        mode & lower limit ($\times 10^{33}$ yr) & $\tan{\beta}
        =10$ & $\tan{\beta}=30$
        \\
            \hline\hline
            $p\to K^+\overline{\nu}$ & 2.3 & 2.3 & 3.5\\
            $p\to K^0\mu^+$ & 1.3 & 2.3 & 1.6\\
            $p\to K^0 e^+$ & 1.0 & * & *\\
            $p\to \pi^0 e^+$ & 10.1 & * & *\\
            $p\to \pi^0 \mu^+$ & 6.6 & 9.8 & 6.6\\
            $p\to \pi^+ \overline{\nu}$ & 0.025 & 1.7 & 2.7\\
            \hline\hline
        \end{tabular}
    \end{center}
    \caption{The predictions for the upper limits on the partial
    lifetimes of various proton decay modes for the new mass fit in our
    model for $m_{1/2}=200$ GeV and $x_1=0.1$. The most stringent constraint is
    from $p\to \pi^0\mu^+$ mode, and hence, The squark mass has been chosen
    to be 1.4 (1.2) TeV for $\tan\beta=10~(30)$ so as to just
    satisfy the most stringent bound. Note that in
    this case, the model does not have any predictions for the decay modes 
    $p\to K^0e^+$
    and $p\to \pi^0 e^+$, because the $C$ coefficients for both these
    modes involve products of (1,1) elements of the Yukawa coupling
    matrices, and by construction, these elements are zero for all the three
    coupling matrices; hence these modes have vanishing decay rates.}
    \label{tab:modb}
\end{table}
The model predictions for $x_1=0.1$ for various decay modes are
given in Table~\ref{tab:modb}. We note that the observation of
one of the decay modes in the last two columns of Table~\ref{tab:modb} 
at a given
rate will fix $x_1$ and the rates for remaining modes (the ones
without stars) are then predicted and should provide a test of
this model.
 It should also be noted here that within the mSUGRA
framework at low $\tan\beta$, Tevatron has put a lower limit of
375 GeV for the squark mass based on an integrated luminosity of 1
fb$^{-1}$. We expect our predicted lower bound on the squark mass
which is of order 1 TeV to be testable at higher luminosities
within the reach of LHC.
\section{Effect of $R$-parity breaking}
So far we assumed matter parity so that there is no $R$-parity
violating terms in the superpotential (i.e. $W'~=~0$). In this
section we discuss the implications for relaxing this assumption on
proton life time. This is an interesting exercise in view of the
fact that in MSSM embedding into $SU(5)$, relaxing $R$-parity (or
matter parity) conservation leads to new contributions to baryon
number violation with arbitrary strength, so that in principle,
such models are not viable without matter parity assumption. We
would like to study in this section the situation in the case of
our $SO(10)$ model.

The most general $R$-parity violating interactions upto dimension-5
operators in our model are the following:
\begin{eqnarray}
W'=M'_a\psi_a \bar{\psi}_H+ \lambda \psi_a \psi_H H+
\frac{\lambda_{abc}}{M_{Pl}}\psi_a\psi_b\psi_c\psi_H+S_aS_bS_c
+\mu^{\prime 2}S_a
\end{eqnarray}
where $\psi_{a,b,c}$ denote matter spinors and $\psi_H$ and
$\bar{\psi}_H$ are Higgs spinor fields. Before proceeding to
discuss their implications, note that $M'_a$ must be of order
TeV otherwise the right handed neutrino field would decouple from
the low energy sector and break the gauge multiplet required to
implement inverse seesaw. There are the following classes of
$R$-parity violating operators that follow from this in conjunction
with the $W_m+W_{SB}$ at the TeV scale:
\begin{equation}
W'({\rm TeV})=M'_aL^c_a\bar\chi^c+\lambda L\Phi
\chi^c+\frac{\lambda_{abc}}{M_{Pl}}{\chi}^c
 \left[Q^c_aQ^c_bQ^c_c+L_aQ_bQ^c_c+L^c_aL_bL_c
       +\cdots \right]
\end{equation}
Note that the first three terms within the square bracket, after
$B-L$ breaking, give rise to the familiar MSSM $R$-parity breaking
terms with however couplings determined to be of order
$\frac{v_{BL}}{M_{Pl}}$ which is of order $10^{-15}$. Hence their
contribution to proton decay is negligible. Note this would not be
the case with $SO(10)$ models where $B-L$ symmetry is broken at the
GUT scale.
\section{Conclusion}
In summary, we have discussed proton decay as well as electroweak
symmetry breaking in a new class of recently proposed $SO(10)$
models with TeV scale $W_R$. We showed in an earlier paper that
the model explains small neutrino masses via the inverse seesaw
mechanism and has the feature of gauge coupling unification. The
right-handed neutrinos in this model are almost Dirac type
(pseudo-Dirac) with masses also in the TeV range making them (as
well as the $W_R$ and $Z'$ bosons)  accessible at the Large Hadron
Collider. The collider signals are different from the case with
Majorana right handed neutrinos of conventional type I seesaw. We
have explored two classes of fermion mass fits in these models. In
both the cases, all the Yukawa couplings entering the dimension-5
proton decay operators are fixed within certain assumptions by
charged fermion mass fits, thereby leading to definite
expectations for the partial lifetimes of various proton decay
modes. We find that it is possible to satisfy the current
experimental lower limits on the lifetimes with a wino mass of
100-200 GeV and squark and slepton masses of order TeV. More
specifically, to satisfy the most stringent bound coming from the
$p\to K^+\overline{\nu}$ decay mode, we need to have a lower limit
of 1.2 (2.1) TeV on the squark  masses in the case of model (A)
for $\tan\beta=10(30)$ and similar lower bounds for model (B) for
a given {\bf 10}-Higgs mixing, assuming the universality of squark
and slepton masses, as in a typical mSUGRA type scenario. Thus,
discovery of squarks at LHC can throw light on the validity of
these models. It is also worth pointing out that the choice of
$SO(10)$ multiplets in this class of models is derivable from
fermionic string compactification.
\begin{acknowledgments}
This work is supported by the US National Science Foundation under
grant No. PHY-0652363. We like to thank M. K. Parida for some
discussions.
\end{acknowledgments}
\appendix
\section{RGEs for soft SUSY-breaking masses in SUSYLR model}
Assuming $R$-parity conservation and the trilinear couplings $A$'s
and $Y$'s  in the superpotential and soft breaking Lagrangian
given by Eqs.~(\ref{eq:superpotential}) and (\ref{eq:softl}) to be
zero, the soft breaking mass RGEs at one-loop level are given
by~\cite{nick}
\begin{eqnarray}
    16\pi^2\frac{d}{dt}m^2_Q &=&
    2m^2_Qh_ah_a^\dag+ h_a\left(2h_a^\dag m_Q^2+4m^2_{Q^c}h_a^\dag+4m^2_{
    \Phi_{ab}}h_b^\dag\right)\nonumber\\
    && -\frac{1}{3}M_1M_1^\dag g_1^2
    - 6M_{2L}M_{2L}^\dag g_{2L}^2 - \frac{32}{3}M_3M_3^\dag g_3^2
    +\frac{1}{8}g_1^2S_2,\\
    16\pi^2\frac{d}{dt}m^2_{Q^c} &=&
    2m^2_{Q^c}h_a^\dag h_a+h_a^\dag\left(2h_am^2_{Q^c}+
    4m^2_Q h_a+4h_bm^2_{\Phi_{ba}}\right)\nonumber\\
    && -\frac{1}{3}M_1M_1^\dag g_1^2
    - 6M_{2R}M_{2R}^\dag g_{2R}^2 - \frac{32}{3}M_3M_3^\dag g_3^2
    -\frac{1}{8}g_1^2S_2,\label{eq:rg1}\\
    16\pi^2\frac{d}{dt}m^2_L &=&
    2m^2_Lh_a'h_a'^\dag+h_a'\left(2h_a'^\dag m^2_L+4m^2_{L^c}h_a'^\dag+
    4m^2_{\Phi_{ab}}h_b'^\dag\right)\nonumber\\
    &&  -3M_1M_1^\dag g_1^2
     - 6M_{2L}M_{2L}^\dag g_{2L}^2 -\frac{3}{8}g_1^2S_2,\\
     16\pi^2\frac{d}{dt}m^2_{L^c} &=&
     2m^2_{L^c}h_a'^\dag h_a'+h_a'^\dag\left(2h_a'm^2_{L^c}+4m^2_Lh_a'+
     4h_b'm^2_{\Phi_{ba}}\right)\nonumber\\
     && + 2\mu^{\alpha^*}_{L^c}\left[m^2_{L^c}\mu^\alpha_{L^c}+m^2_{
     \bar\chi^c}\mu^\alpha_{L^c}+\mu^\beta_{L^c}\left(m^2_S\right)_
     {\beta\alpha}\right]\nonumber\\
      && -3M_1M_1^\dag g_1^2
    - 6M_{2R}M_{2R}^\dag g_{2R}^2
    +\frac{3}{8}g_1^2S_2,\\
    16\pi^2\frac{d}{dt}m_{\bar\chi^c}^2 &=&
    2\mu^{\alpha^*}_{L^c}\left[m^2_{L^c}\mu^\alpha_{L^c}+m^2_{
     \bar\chi^c}\mu^\alpha_{L^c}+\mu^\beta_{L^c}\left(m^2_S\right)_
     {\beta\alpha}\right]\nonumber\\
    &&
    -3M_1M_1^\dag g_1^2 - 6M_{2R}M_{2R}^\dag g_{2R}^2 -
    \frac{3}{8}g_1^2 S_2,\\
    16\pi^2\frac{d}{dt}m_{\chi^c}^2 &=&
    -3M_1M_1^\dag g_1^2 - 6M_{2R}M_{2R}^\dag g_{2R}^2 +
    \frac{3}{8}g_1^2 S_2,\\
    16\pi^2\frac{d}{dt}\left(m^2_S\right)^{\alpha\beta} &=&
    4\mu^{\alpha^*}_{L^c}\mu^\beta_{L^c}\left(m^2_{\bar\chi^c}+m^2_{L^c}\right)
    + 2 \mu^{\alpha^*}_{L^c}\mu^\rho\left(m_S^2\right)_\rho^\beta,\\
    16\pi^2\frac{d}{dt}m^2_{\Phi_{ab}} &=& m^2_{\Phi_{ac}}
    {\rm Tr}\left(3h^\dag_c h_b +h'^\dag_c h'_b\right)
    +{\rm Tr}\left(3h^\dag_a h_c+h'^\dag_a h'_c\right)
    m^2_{\Phi_{cb}}\nonumber\\
    &&
    + {\rm Tr}\left(6h^\dag_a h_b m^2_{Q^c}+6h^\dag_a m^2_Q h_b +
    2h'^\dag_a h'_b m^2_{L^c}+ 2h'^\dag_a m^2_L h'_b\right)\nonumber\\
    &&
    + \left(
    -6M_{2L}M_{2L}^\dag g^2_{2L} - 6M_{2R}M_{2R}^\dag g^2_{2R}\right)\delta_{ab}
    \label{eq:rgn}
\end{eqnarray}
where
\begin{eqnarray}
    S_2&\equiv& 4\left[{\rm Tr}\left(m^2_{Q}-m^2_{Q^c}-
    m^2_{L}+m^2_{L^c}
    \right) + \left(m^2_{\chi^c}-
    m^2_{\bar{\chi}^c}\right)
    \right]
\end{eqnarray}
We have ignored the RG running of the coupling $\mu^\alpha_{L^c}$
as these are higher order effects.
\section{Anomalous dimensions of the dimension-5 operator}
Here we present the derivation of the anomalous dimensions of the dimension-5
operators of the $LLLL$ type given by Eq.~(\ref{eq:ol}).
The calculation is straightforward in a supersymmetric gauge
due to the fact that the operator ${\cal O}_L$
is purely chiral (it is an $F$-term), and hence, it follows from
non-renormalization theorems that in a supersymmetric gauge, it
will only have wave function renormalization. Then it is easy to show that
the anomalous dimensions of any purely chiral operator are given by
\begin{equation}
    \gamma_{\cal O} = \sum_{r}C_2(r)
\end{equation}
where $C_2(r)$ is the eigenvalue of the quadratic Casimir operator in the
representation $r$, and the sum runs over all the chiral superfields occurring
in the chiral coupling. As the gauge bosons belong to the adjoint
representation, we have
\begin{eqnarray}
    C_2(r) = \left\{\begin{array}{cc}
        \frac{N^2-1}{2N} & {\rm for}~SU(N)\\
        \frac{1}{4}X^2 & {\rm for}~U(1)_X
    \end{array}\right.
\end{eqnarray}
Thus we have for $SU(3)_c$,
\begin{eqnarray}
    \gamma_{\mathbf 3_c} = 3\times \frac{4}{3} = 4
\end{eqnarray}
as there are three $SU(3)_c$ fields in the $LLLL$ operator
[e.g. $(qq)(\tilde{q}\tilde{l})$]. Similarly, we have
\begin{eqnarray}
    \gamma_{\mathbf 2_{L,R}} &=& 4\times \frac{3}{4} = 3,\\
    \gamma_{\mathbf 1_Y} &=& \frac{1}{4}\left[3\left(\frac{1}{3}\right)^2
    +1\right]\frac{3}{5} = \frac{1}{5},\\
    \gamma_{\mathbf 1_{B-L}} &=& \frac{1}{4}\left[3\left(\frac{1}{3}\right)
    ^2+1\right]\frac{3}{2} = \frac{1}{2}
\end{eqnarray}
Here the factors $\frac{3}{5}$ and $\frac{3}{2}$ are the GUT normalization
factors for $U(1)_Y$ and $U(1)_{B-L}$ respectively.

We note that the same results would have been obtained in a
non-supersymmetric gauge, though the calculation is much more involved.
For instance, the same results were obtained for the MSSM case in a
Wess-Zumino gauge in Ref.~\cite{ibanez}.
\section{The hadronic factors $f(F,D)$}
As noted in Sec. VI, the hadronic factor $f(F,D)$ estimates the chiral
symmetry breaking effects on different final states. The low-energy parameters
$D$ and $F$ are usually chosen to be the same as the analogous parameters
in weak semileptonic decays~\cite{marshak}. Then $D+F=g_A^{(np)} =1.27$ is
the nucleon axial charge, while $D-F=g_A^{(\Sigma^-n)}=0.33-0.34$~\cite{pdg}.
This gives $D=0.8$ and $F=0.47$. Using these constants and the approximations
$m_{u,d}\ll m_s \ll m_p$ as well as $-q^2\ll m_p^2$ where $q_\mu$ is the
momentum transfer (the momentum of the anti-lepton for physical decays), all
the
hadronic matrix elements can be obtained~\cite{lattice}. In Table IV, we list
the results for different decay channels.
\begin{table}[h!]
    \begin{center}
        \begin{tabular}{||c|c|c||}\hline\hline
            Decay mode & $f(F,D)$ &
            $|f(F,D)|^2$ \\ \hline\hline
            $p\to \pi^0 l^+$ &
    $\frac{1}{\sqrt 2}(1+D+F)$ & 2.58 \\
    $p\to \pi^+ \overline{\nu}_l$ &
    $1+D+F$ & 5.15\\
    $p \to K^0 l^+$ & $1-\frac{m_N}{m_B}(D-F)$ & 0.53\\
    $p \to K^+ \overline{\nu}_l$ & $\frac{m_N}{m_B}\frac{2D}{3}$ & 0.19\\
    \hline\hline
        \end{tabular}
    \end{center}
    \caption{The hadronic factors $f(F,D)$ for different proton decay
    modes. Here we have used $m_N=0.94$ GeV for the mass of nucleon and
    $m_B=1.15$ GeV for the average baryon mass
    ($m_B\simeq m_{\Sigma}\simeq m_{\Lambda}$).}
\end{table}

\end{document}